\documentclass[aps,prd,twocolumn, english,reprint, longbibliography, superscriptaddress, breaklinks=true, showkeys, showpacs=false, nofootinbib]{revtex4-2}
\usepackage[T1]{fontenc}
\usepackage[latin9]{inputenc}
\usepackage{color}
\usepackage{babel}
\usepackage{amsmath}
\usepackage{amsthm}
\usepackage{amsfonts}
\usepackage{amssymb}

\usepackage{subfigure}
\usepackage{graphicx}
\usepackage{physics}

\usepackage{hyperref}
\makeatletter

\pdfoutput=1
\hypersetup{colorlinks=true,citecolor=red,linkcolor=blue,urlcolor=blue,filecolor= green, breaklinks=true}
\begin{document}

\title{Regular Kerr black holes: Junction conditions and the matter content across the ring}

\author{Marcos L. W. Basso}
\email[Corresponding author: ]{marcoslwbasso@hotmail.com}
\address{Centro de Ci\^encias Naturais e Humanas, Universidade Federal do ABC, Avenida dos Estados 5001, Santo Andr\'e, S\~ao Paulo, 09210-580, Brazil}

\author{Vilson T. Zanchin}
\address{Centro de Ci\^encias Naturais e Humanas, Universidade Federal do ABC, Avenida dos Estados 5001, Santo Andr\'e, S\~ao Paulo, 09210-580, Brazil}

\begin{abstract}
Regular rotating black holes are usually described by a metric of the Kerr-Schild form with a particular mass function that is chosen to avoid the ring singularity of the Kerr metric and which approaches the Kerr metric at the asymptotic limit.  However, as is well known, even for a class of well-behaved mass functions, the curvature scalars present a discontinuity in the equatorial plane at the ring. This discontinuity has been associated with the presence of a string of matter that joins the interior and exterior regions along the equatorial plane. By using the Darmois-Israel junction conditions, we analyze all four possible combinations of the normal vector orientations on each side of the ring, construct the complete stress-energy momentum tensor of the string, and interpret each resulting solution. We show that, out of the four possibilities, only one of the four models for the string solution at the ring yields the appropriate asymptotic geometry. In such a case, the string bears a fluid with nonzero pressure, but with a vanishing line energy density, and it does not rotate at all. Finally, taking an appropriate metric for the exterior region, we also discuss a different scenario in which the matter source at the ring is a rotating lightlike fluid. 
\end{abstract}


\maketitle

\section{Introduction}
\label{sec:intro}

The Kerr metric~\cite{Kerr67} has been widely studied in the general relativity literature for two main reasons. The first reason is because it describes the final state of rotating matter distributions after the complete gravitational collapse, the results of which being rotating black holes. The second reason is related to the intricate properties of such a metric in the region inside the event horizon, which contains a ring singularity,  causality violations, and closed timelike curves~\cite{Neil}. A potential remedy for these issues involves replacing the problematic interior region of the Kerr geometry with an appropriate matter source. In addition to avoid the singularity, the source must respect reasonable properties such as stationarity, axial symmetry, separability of the Hamilton-Jacobi equations, and constraining the amount of exotic matter that violates the energy conditions in the region inside the event horizon, once such exotic matter is not observed in astrophysical scenarios~\cite{Simpson}.

In Ref.~\cite{Gurses}, the authors demonstrated that, in the context of Kerr-Schild spacetimes~\cite{Schild}, a Lorentzian coordinate system can be selected in such a way that it leads to the linearization of the Einstein field equations. As an application of this result, the authors showed that complex translations can be applied in order to obtain new solutions for spinning systems, including new anisotropic interior metrics that may be matched to the Kerr metric on an oblate spheroid. This procedure paved the way for constructing rotating counterparts of static and spherically symmetric regular black hole solutions resembling the Kerr metric in the asymptotic region, but with the constant mass parameter replaced by a mass function $m(r)$, so modifying the Kerr geometry close to the central ring. We refer to this kind of rotating metric as the G\"urses-G\"ursey metric. Subsequent explorations have been devoted to understanding the source generating the rotating metric~\cite{Burinskii, Gondolo} and a substantial body of work on regular rotating objects has emerged~\cite{Spallucci, Modesto, Bambi, Mustapha, Tosh14, Saa, Dymn15, Ghosh2015, Tosh17, Hernandez-Pastora:2017fmg, Torres, Gosh, Mazza, Franzin, Masa22, Maeda, Simpson, Brustein, Dymn2023,  Ramon, Casadio:2023iqt, BZ2024-paper1, BZ2024-paper2}.

In particular, Torres~\cite{Torres} and Maeda~\cite{Maeda} have shown that, in contrast to the Kerr metric, if the mass function $m(r)$ can be expanded around $r = 0$ as $m(r) \approx m_0 r^{3 + \alpha}$ with $m_0 \neq 0$ and $\alpha \ge 0$, the ring $\mathcal{S}\! :\! (r = 0,\, \theta = \pi/2)$ corresponds to a conical singularity when extended to the region $r<0$, and not to a scalar polynomial curvature singularity. Moreover, if the extension to the region $r<0$ is not performed, Torres~\cite{Ramon23} showed that the ring is devoid of conical singularities as well. 

Although these different types of singularity can be avoided in the ring for the G\"urses-G\"ursey rotating metric \cite{Gurses}, it is well known that the curvature scalars are not well-defined there, once they present a finite jump which depends on the path taken when approaching the ring for $\alpha  = 0$~\cite{Spallucci, Modesto, Torres, Maeda, Masa22, BZ2024-paper1, BZ2024-paper2}. On the other hand, if $\alpha > 0$ the curvature scalars are continuous and does not depend on the path taken when approaching the ring~\cite{Ramon23}. Interestingly, the curvature scalars are also well-defined and continuous for the mass function with exponential suppression, that is, $m(r) = m_0 e^{-l/r}$~\cite{Ghosh2015, Simpson}, where $l$ is a parameter that quantifies the deviation from the Kerr geometry. 
These kinds of regular Kerr geometries with an exponential mass function represent rotating regular black holes with an asymptotically Minkowski core~\cite{Simpson}.

The feature exhibited by the G\"urses-G\"ursey metric for $\alpha = 0$ at the ring has been interpreted by Smailagic and Spallucci~\cite{Spallucci} as a string replacing the ring singularity observed in the Kerr solution. More specifically, since the limit $r \to 0$ for $\theta = \pi/2$ corresponds to the curvature of a de Sitter-like geometry, while the limit $r \to 0$ for $\theta \neq \pi/2$ corresponds to the curvature of the Minkowski geometry~\cite{Torres}, the authors glued a section of the Minkowski spacetime with a de Sitter-like spacetime at the ring and, 
for unconventional orientations of the normal vectors, interpreted the discontinuity of the extrinsic curvature as due to the presence of a string of matter at the ring.

The finite jump presented by the curvature scalars at the Kerr ring deserves further investigation and, therefore, motivated by the previous contributions mentioned above, in this work we study in detail the proposal of replacing the ring by a string and explore all possible choices of the signs of the normal vectors at the ring. Let us remind that the normal vector of a thin shell plays an important role, since it determines extrinsic curvature of the shell regarding the spacetime in which the shell is immersed~\cite{Israel66}. In fact, to uncover all possible thin shell solutions, it is necessary to take into account that the normal to a shell can have two distinct orientations relative to the center of the coordinates of both the interior and exterior spacetimes (or spacetime regions), even though the normal vector is fixed to point always from a given spacetime to the other~\cite{Katz, Dalia, Lemos22}. Moreover, we also show that the string that replaces the Kerr ring in the scenario presented in Ref.~\cite{Spallucci} is actually static. To address this issue, we construct a scenario in which the string is indeed rotating.

The present work is organized as follows. In Sec.~\ref{sec:II}, we review the regular Kerr geometry and explore its main properties, in particular, the behavior of the relevant curvature scalars and other quantities close to the ring. We also present the metrics for the regions inside, outside, and the ring. The Darmois-Israel junction conditions are briefly described in Sec.~\ref{Sec:III}. The matching at the ring using the same spacetime construction as in Ref.~\cite{Spallucci} is done in Sec.~\ref{sec:nonrot-ring}, where we uncover all possible thin-shell (string) solutions and show that only one of the four possibilities yields the appropriate geometry and show that neither of them corresponds to a rotating string. In Sec.~\ref{Sec:V}, we present a new spacetime construction in which the string is indeed rotating, and we analyze the four possible thin-shell solutions. Our final comments are made in Sec.~\ref{sec:conc}.

\section{The regular rotating geometry}
\label{sec:II}
\subsection{The spacetime metric}
\label{sec:stmetric}
 
In Boyer-Lindquist coordinates, the metric of interest here reads
\begin{equation}
\begin{aligned} 
   ds^2  = & -\left(1 - \frac{2 r\, m(r) }{\Sigma(r,\theta)} \right) dt^2 + \frac{\Sigma(r,\theta)}{\Delta(r)} dr^2   \\ & + \Sigma(r,\theta) d\theta^2 - \frac{4 r\, m(r)  a \sin^2 \theta}{\Sigma} dt\, d\varphi  \\
    & + \left(r^2 + a^2 + \frac{2 r\, m(r) a^2 \sin^2 \theta}{\Sigma(r,\theta)} \right) \sin^2\theta\, d\varphi^2,  
\end{aligned} \label{eq:ggmetric}
\end{equation}
where $m(r)$ stands for the mass function, $a$ indicates the rotation parameter, which we assume to be positive, and $\Sigma(r,\theta)$ and $\Delta(r)$ are defined by 
\begin{align}
& \Sigma(r,\theta) = r^2+ a^2\cos^2\theta, \label{eq:sigmaa}\\
& \Delta(r) = r^2 + a^2 - 2r m(r), \label{eq:deltfun}
\end{align}
respectively.

In order to make our analysis as general as possible, we do not choose any specific mass function $m(r)$. The only imposition on $m(r)$ is that it can be expanded around $r = 0$ as
\begin{equation}
m(r) = m_0 r^{3 + \alpha}, \label{eq:m(r)}
\end{equation}
with $m_0 \neq 0$ and $\alpha \geq 0$. This assumption guarantees that the desired properties of metric \eqref{eq:ggmetric} as summarized in Sec.~\ref{sec:intro} remain valid, it assures particularly that the curvature scalars do not diverge in the limit $r\to 0$. This is the main reason why such an asymptotic behavior of the mass function is present in most of the works cited here, especially the work by Smailagic and Spallucci~\cite{Spallucci}.

The Ricci scalar for the metric \eqref{eq:ggmetric} is given by
\begin{align}
    \mathcal{R} = \frac{2}{\Sigma(r,\theta)}\big[2m'(r) + r\,m''(r)\big], \label{eq:ricci}
\end{align}
with the primes indicating differentiation with respect to the coordinate $r$. Equations \eqref{eq:m(r)} and \eqref{eq:ricci} tell us that, around $r = 0$, the Ricci scalar is given approximately by 
\begin{align}
    \mathcal{R} =
     2\left(3+\alpha\right)\left(4+\alpha\right)\frac{m_0\,r^{2 + \alpha}}{\Sigma(r,\theta)}. \label{eq:ricci2}
\end{align}
Hence, for $\theta \neq \pi/2$, the limit $r \to 0$ gives $\mathcal{R} = 0$ for all $\alpha\geq 0$.
In turn, for $\theta = \pi/2$, Eq.~\eqref{eq:ricci2} furnishes $2\left(3+\alpha\right)\left(4+\alpha\right)m_0r^{\alpha}$, and therefore, the limit $r \to 0$ gives $\mathcal{R} = 0$ for $\alpha> 0$, but it gives $\mathcal R = 24 m_0$ for $\alpha = 0$. In conclusion, if the mass function is of the form $m(r)\sim r^{3 + \alpha}$ close to $r\to 0$, then the Ricci scalar is well-defined for all values of $\alpha>0$. However, for $\alpha=0$, in the limit to the ring ($r\to 0, \theta\to \pi/2$) the Ricci scalar is a double-valued function, i.e., $\mathcal{R}(r=0,\theta= \pi/2+ \epsilon) = 0$ but $\mathcal{R}(r=0,\theta= \pi/2) = 24m_0$,  and hence it is not well-defined there. The same happens also with the other independent curvature scalars~\cite{Torres}.

For instance, the Ketschmann scalar results as
\begin{equation}
\begin{split}
      \mathcal{K}=& \dfrac{4\,m_0^2 r^{2(\alpha +2)}}{\Sigma^2(r,\theta)}\bigg[\dfrac{384\,  r^8}{\Sigma^{4}(r,\theta)} -\dfrac{192\left(\alpha+6\right)\, r^{6}}{\Sigma^3(r,\theta)} \\ &
   +\dfrac{8\,r^{4}}{\Sigma^2(r,\theta)}\big[27 + 2 (\alpha+ 3) (3\alpha+ 22)\big]
   \\
    &-\dfrac{4\, r^{2}}{\Sigma(r,\theta)}\big[(\alpha+4)\big(2\alpha(\alpha+9) +39 \big)\big] \\ 
   & + {4(\alpha+3)^2(\alpha+4)^2}\bigg] .
    \end{split}
\end{equation}

The last equation shows that the behavior of the Kretschmann scalar in the region $r\to 0$ is similar to the Ricci scalar.
Assuming $\theta\neq\pi/2 $, the limit $r\to 0 $ gives ${\cal K} =0$ for $\alpha\geq 0$. However, for $\theta=\pi/2$ we get ${\cal K}(r,\pi/2)_{r\to 0} = 4\left( \alpha^4+6\alpha^3+17\alpha^2 +28\alpha+24\right) m_0^2\, r^{2\alpha}$, which gives zero for $\alpha >0$, but gives ${\cal K}= 96 m_0^2$ for $\alpha =0$. In conclusion, we verify that for $\alpha=0$ the Kretschmann scalar is double-valued in the Kerr ring, similar to the Ricci scalar, an undesired feature that deserves further investigation. 

The source that generates the G\"urses-G\"ursey geometry is an anistropic fluid~\cite{Gurses, Burinskii, Gondolo}, whose energy-momentum tensor may be cast into the form 
\begin{align}
    T^{\mu \nu} = \rho_m\, e_0^{\ \mu}e_0^{\ \nu} + \mathfrak p_1 e_1^{\ \mu}e_1^{\ \nu} + \mathfrak p_{2} e_2^{\ \mu}e_2^{\ \nu} + \mathfrak p_{3} e_3^{\ \mu}e_3^{\ \nu},\label{eq:ggt}
\end{align}
where $\rho_m$ is the energy density, $\mathfrak p_i$ ($i=1,\,3,\,3)$ are the pressures, and 
\begin{align}
    & e_0^{\ \mu} = \frac{1}{\sqrt{\pm \Delta(r) \Sigma}}\Big(\left[r^2 +a^2\right]\delta^{\mu}_{\ t} + a\, \delta^{\mu}_{\ \varphi}\Big), \nonumber \\
    & e_1^{\ \mu} = \sqrt{\frac{\pm \Delta(r)}{\Sigma}} \delta^{\mu}_{\ r}, \ \ e_2^{\ \mu} = \frac{1}{\sqrt{\Sigma}} \delta^{\mu}_{\ \theta}, \label{eq:rtetrad} \\
    & e_3^{\ \mu} =  \frac{1}{\sqrt{\Sigma}\sin \theta}\big[a \sin^2 \theta\delta^{\mu}_{\ t} + \delta^{\mu}_{\ \varphi}\big],\nonumber
\end{align}
is the well known Carter's orthonormal frame, which diagonalizes the energy-momentum tensor according to Eq.~\eqref{eq:ggt}. For regular black holes, the plus sign applies to the regions outside the event horizon and inside the Cauchy horizon, with $e_0^{\ \mu}$ being a unit timelike vector that describes the four-velocity of the fluid. In turn, the minus sign applies to the region between the event horizon and the Cauchy horizon, with the vectors $e_0^{\ \mu}$ and $e_1^{\ \mu}$ switching roles.

The total energy density $\rho_m$, the radial pressure $\mathfrak p_1$, and the tangential pressures $\mathfrak p_2$ and $\mathfrak p_3$ are given by~\cite{Burinskii}
\begin{align}
    & \rho_m(r,\theta) = -\mathfrak p_1(r,\theta) = \frac{r^2 m'(r)}{4 \pi \Sigma^2}, \label{eq:rpressa}\\
    & \mathfrak p_{2}(r,\theta) = \mathfrak p_{3}(r,\theta) = \frac{r^2 m'(r)}{4 \pi \Sigma^2} - \frac{1}{8 \pi \Sigma}\Big(r\,m(r)\Big)'' \label{eq:tpressa},
\end{align} 
with the prime indicating derivative with respect to the coordinate $r$. The above relations for the fluid quantities hold for any mass function $m(r)$, but we are interested in the cases where the mass function is such that $m(r)\sim r^3$ as $r\to 0$.

The timelike vector $e_0^{\ \mu}$ is the four-velocity of a stationary fluid that rotates with the angular velocity
\begin{align}
   \Omega(r) = \frac{e_0^{\ \varphi}}{e_0^{\ t}} = \frac{a}{r^2 + a^2}. \label{eq:GGveloc} 
\end{align} 

Another interesting property of the regular Kerr metric is the so-called frame dragging. The frame-dragging velocity $\omega$ is defined by considering an observer that, starting from infinity with zero angular momentum, falls freely toward the region defined by $r = 0$. Such conditions define the so-called zero angular momentum observers. In the Boyer-Lindquist coordinates, one has 
\begin{align}
\omega(r,\theta) = - \frac{g_{\varphi t} }{g_{\varphi \varphi}}.
\label{eq:dragg(r)}
\end{align}
where 
\begin{align}
g_{\varphi t} & = -\frac{2r\,m(r)\, a \sin^2 \theta}{\Sigma}, \label{eq:gtphi}\\
g_{\varphi \varphi} & = \left( r^2 + a^2 + \frac{2r\, m(r)\, a^2 \sin^2 \theta}{\Sigma}\right) \sin^2 \theta.\label{eq:gphphi}
\end{align}

The frame-dragging velocity \eqref{eq:dragg(r)} is well-defined everywhere, including the ring ($r=0$, $\theta=\pi/2$) where it vanishes.
In the equatorial plane, $\theta=\pi/2$, it gives $\omega(r,\pi/2) = 2m(r)\, r^2 a\big/\left[r^2\left(r^2 + a^2\right) +2m(r)\, r\,a^2\right]$. So, by considering a mass function that close to $r=0$ is of the form \eqref{eq:m(r)},  the limit $r\to 0$ yields 
\begin{equation}
    \lim_{r\to 0}\omega(r,\pi/2) = 0.
\end{equation}
This important result is different from the singular Kerr metric in which case one has $  \lim_{r\to0} \omega(r,\pi/2) = 1/a$.
For $\theta\neq\pi/2$, the situation here is the same as for the singular Kerr geometry, namely,
\begin{equation}
       \lim_{r\to 0}\omega(r,\theta\neq \pi/2) = 0 , 
\end{equation}
so that the frame-dragging velocity is well-defined everywhere in the regular Kerr-like geometry.

As just mentioned above, the singular locus of the Kerr metric parameterized by ($r=0, \, \theta=\pi/2$) in the Boyer-Lindquist coordinates can be regularized by replacing the constant mass parameter of the Kerr metric with a power-law mass function of the form \eqref{eq:m(r)}, but just for $\alpha >0$. For $\alpha < 0$, the locus is still a curvature singularity as in the Kerr metric. 
The case $\alpha =0$ is special since the curvature scalars are all finite but present a discontinuity at that locus. This feature motivated the authors of Ref.~\cite{Spallucci} to introduce a string replacing the ring singularity of the Kerr spacetime. Following this idea, in the next section, we shall explore the proposal by Smailagic and Spallucci in more detail.

\subsection{Inside the ring: The disk}
\label{sec:disk}

The limit $r \to 0$ for $\theta \neq \pi/2$ of the metric \eqref{eq:ggmetric} corresponds to a disk $\mathcal{D}$ of radius $a$ that can be parameterized by the polar coordinates $(\rho,\, \varphi)$ so that $0\leq \rho\le a$ and $0\leq\varphi\leq 2\pi$. The coordinate $\rho$ is given in terms of the Boyer-Lindquist coordinates by 
\begin{equation}
\rho = a \sin \theta, \label{eq:rho}
\end{equation}
while $\varphi$ is the azimuth angle. At the disk $\mathcal{D}\! : \! (r = 0, 0\leq \theta < \pi/2)$, the metric \eqref{eq:ggmetric} reduces to
\begin{align}
    ds^2_{i} = -dt^2 + d\rho^2 + \rho^2 d \varphi^2, \label{eq:Dmetric}
\end{align}
which is the flat metric, indicating a Minkowskian region with vanishing curvature scalars.

\subsection{The ring}
\label{sec:ring}

The ring $\mathcal{S}$ is the boundary of the disk defined in the last paragraph, i.e., it may be parameterized by $(\rho=a, 0\leq\varphi\leq 2\pi)$.
At the ring $\mathcal{S}\!:\!(r=0, \theta = \pi/2)$, metric \eqref{eq:ggmetric} reduces to 
\begin{align}
    ds^2_{\mathcal{S}} = - dt^2 + a^2 d \varphi^2, \label{eq:Smetric}
\end{align}
where, from Eq.~\eqref{eq:rho}, the assumptions $r=0$ and $\theta=\pi/2$ imply $\rho=a$.

\subsection{Outside the ring: The equatorial plane}  
\label{sec:equatorial}

For the present analysis it is useful to have the explicit form of metric \eqref{eq:ggmetric} at the equatorial plane, i.e., for $\theta = \pi/2$. 
First we define a new radial coordinate $\varrho$ by
\begin{equation}
  \varrho  = \sqrt{r^2 + a^2}, \label{eq:varrho}
\end{equation}
 whose range is $\varrho \in [a, \infty)$.
With this, at $\theta= \pi/2$, metric~\eqref{eq:ggmetric} assumes the form 
\begin{equation}
\begin{aligned}
ds^2_{\scriptscriptstyle{E}} =  & -\left(1 - \frac{2 m(\varrho) }{\sqrt{\varrho^2 - a^2}} \right) dt^2 - \frac{4 m(\varrho)  a }{\sqrt{\varrho^2 - a^2}} dt\, d\varphi  \\ & + \frac{\varrho^2}{\Delta(\varrho)} d\varrho^2 
     + \left(\varrho^2 + \frac{2 m(\varrho) a^2}{\sqrt{\varrho^2 - a^2}} \right) d\varphi^2, 
\end{aligned}  \label{eq:Emetric}
\end{equation} 
where $\Delta(\varrho)= \varrho^2 - 2m(\varrho)\sqrt{\varrho^2 - a^2}$.

Approaching the ring, but remaining outside, i.e., in the region $\mathcal{A}\! :\! (r \to 0^+,\, \theta = \pi/2)$, 
the mass function $m(\varrho)$ is given approximately by the expression in Eq.~\eqref{eq:m(r)}, namely,
\begin{equation}
    m(\varrho) = m_0 (\varrho^2 - a^2)^{3/2}, \label{eq:rhoapprx}
\end{equation}
and, to the fist order in $\rho^2-a^2$, the metric \eqref{eq:Emetric} reduces to 
\begin{align}
   ds^2_{e} = &\,  - \left[1 - 2m_0\left(\varrho^2 - a^2\right)\right] dt^2 -4m_0a\left(\varrho^2 - a^2\right)dtd\varphi \nonumber \\
&\; + d \varrho^2 + \left[\varrho^2 +  2m_0a^2\left(\varrho^2 - a^2\right)\right]  d \varphi^2, \label{eq:Ametric0}
\end{align}
with $\varrho \in [a,\, a + \varepsilon)$ for $\varepsilon > 0$. This is the metric assumed to describe the exterior region of the ring. Note that \eqref{eq:Ametric0} is the general rotating Kerr-like metric \eqref{eq:ggmetric} with $\theta=\pi/2$ and $m(r) = m_0r^3$, i.e., it is the complete rotating metric \eqref{eq:Emetric} with $m(\varrho)= m_0\left(\varrho^2-a^2\right)^{3/2}$, without any further approximation.

In Ref.~\cite{Spallucci}, the chosen exterior metric is
\begin{align}
&    ds^2_{e} = - \left[1 - 2m_0\left(\varrho^2 - a^2\right)\right] dt^2 + d \varrho^2 + \varrho^2 d \varphi^2. \label{eq:Ametric}
\end{align} 

In comparison to the rotating metric \eqref{eq:Ametric0}, the metric \eqref{eq:Ametric} neglects the terms containing $m_0a\left(\varrho^2- a^2\right)$ and $m_0a^2\left(\varrho^2- a^2\right)$. As a consequence, the $g_{t\varphi}$ coefficient vanishes and then the result is a static (nonrotating) metric.  
In fact, the metric \eqref{eq:Ametric} can be considered an approximation for the static de Sitter (dS) metric, or for the anti-de Sitter metric, depending on the sign of $m_0$.
The authors in \cite{Spallucci} have chosen $m_0 > 0$, which leads to a de Sitter-like metric. In turn, metric \eqref{eq:Ametric0} is appropriate for the exterior metric, since it is exactly the rotating metric \eqref{eq:ggmetric} in the equatorial with a properly chosen mass function.

It is worth mentioning that most regular black holes described in the literature present a de Sitter-type core, which corresponds to $m_0 > 0$. Even though we are not aware of any specific work that constructs regular black holes with anti-de Sitter-type cores, i.e., with $m_0 < 0$, in Ref.~\cite{Bargueno} the author shows that, among other kinds of objects, it is possible to construct such regular black holes.

\section{The matching at the ring}
\label{Sec:III}

Although the Darmois-Israel formalism (DIF)~\cite{Israel66} for matching different metrics is well known, we review the main aspects of such a formalism here in order to set up notation and present the important quantities adapted to the present problem.

Let us then refer to the interior disk $\mathcal{D}$ with the metric given by Eq.~\eqref{eq:Dmetric} as the spacetime $(\mathcal{M}_i,\, g_i)$, and to the equatorial region outside the ring $\mathcal{A}$ with the metric given by Eq.~\eqref{eq:Ametric} in a first study and by \eqref{eq:Ametric0} in a second study, as the spacetime $(\mathcal{M}_e,\,g_e)$.  These two spacetimes are to be matched at the ring $\mathcal{S}$. Let $\xi^a = (t, \varphi)$ be the intrinsic coordinates of the ring $\mathcal{S}$, whose metric is given by Eq.~\eqref{eq:Smetric}. Without loss of generality, we have already identified the timelike coordinate $t$ and the azimuthal coordinate $\varphi$ in the three regions $\mathcal{M}_{i}$, $\mathcal{M}_e$, and $\mathcal{S}$.

The DIF deals with the first and second fundamental forms on $\mathcal{S}$, $h_{ab}$ and $K_{ab}$, respectively. Such quantities may be defined in terms of geometric quantities in the spacetimes $\mathcal{M}_{i}$ and $\mathcal{M}_{e}$ by relations of the form $h_{ab} = e_a^{\ \mu} e_b^{\ \nu}g_{\mu \nu}$ and $K_{ab} = e_a^{\ \mu} e_b^{\ \nu} \nabla_{\nu} n_{\mu}$,  where $e_{a}^{\ \mu} = \partial x^{\mu} / \partial \xi^a$, $n_{\mu}$ is the spacelike unit vector normal to the ring $\mathcal{S}$, and $\nabla_{\nu}$ stands for the covariant derivative compatible with the Lorentzian metric. There is one such relation with a corresponding quantity for each one of the spacetime regions $\mathcal{M}_{i}$ and $\mathcal{M}_{e}$. For example, the unit normal vector $n_\mu$ has two representations: $n_{(i)u}$ is the unit normal vector as seen from the point of view of the internal section $\mathcal{M}_{i}$, while $n_{(e)\mu}$ is the unit normal vector as seen from the point of view of the external section $\mathcal{M}_{e}$.

Moreover, in the DIF, it is necessary to choose the orientation of the unit normal vector to the boundary between the two spacetime regions ($n_{(i)\mu}$ and $n_{(e)\mu}$), which we choose to point from $\mathcal{M}_i$ to $\mathcal{M}_e$. The normal to a closed thin shell may have two distinct behaviors relative to the center of the coordinates in each spacetime, i.e., its orientation regarding the increasing or decreasing of the radial coordinate of each spacetime is not fixed a priori \cite{Lemos22}. This implies that different sections of the interior and exterior spacetimes bounded by the ring can be glued to each other. Understanding this fact is essential for uncovering all possible thin shell solutions for the present problem.

The first boundary condition in the DIF is the continuity of the first fundamental form across the boundary $\mathcal{S}$, i.e., $[h_{ab}] = 0$. Here we employ the standard notation $[Q] \equiv Q\big|^e_{\mathcal{S}} - Q\big|^i_{\mathcal{S}}$, with $Q\big|^{e}_{\mathcal{S}}$ denoting any given quantity $Q$ evaluated on $\mathcal{S}$ from the perspective of region $\mathcal{M}_{e}$, and similarly for $Q\big|^{i}_{\mathcal{S}}$. It is easy to see that, in the present case, the induced metric $h_{ab}$ on $\mathcal{S}$ as perceived from each spacetime region is the same and, with our choice of coordinates, is given by Eq.~\eqref{eq:Smetric}.

The second boundary condition establishes the relation between the possible discontinuity of the second fundamental form $K_{ab}$ across the boundary $\mathcal{S}$ and the energy-momentum tensor (EMT) of the boundary layer. I.e., if $[K_{ab}]\neq 0$ then a thin shell at $\mathcal{S}$ is needed, and the respective energy-momentum tensor $S_{ab}$ is given by
\begin{align}
    8 \pi S_{ab} = -[K_{ab}] + h_{ab} [K], \label{eq:EMT}
\end{align}
where $K = h^{ab}K_{ab}$.

\section{Delving into the middle-Kerr: The nonrotating string}

\label{sec:nonrot-ring}

\subsection{The interior region} \label{sec:interior1}

In the spacetime $(\mathcal{M}_i,\, g_i)$ interior to the ring,  the metric is 
\eqref{eq:Dmetric}  and the spacelike unit normal vector to the boundary $\mathcal{S}$ as seen from $\mathcal{M}_i$ is given by
\begin{equation}
    n_{(i)\mu} = \epsilon_i \delta_{\mu}^{\rho},  \label{eq:normal-}
\end{equation}
 where $\epsilon_i$ is a normalization factor. The normalization condition $g^{\mu\nu}_{i} n_{(i)\mu}n_{(i)\nu} = 1$ with the metric \eqref{eq:Smetric} implies $\epsilon_i = \pm 1$. The sign $\epsilon_i$ is the quantity that determines the relative orientation of $n_{(i)\mu}$ in relation to the coordinate $\rho$, as one can see from the scalar product between $n_{(i)\mu}$ and the gradient of $\rho$, i.e., $\nabla_{\mu} \rho$. For $\epsilon_i = +1$, we have $g^{\mu\nu}_{i}n_{(i)\mu}(\nabla_{\nu} \rho) = +1 > 0$, which implies that the normal points in the direction of increasing $\rho$ as seen from $\mathcal{M}_i$. On the other hand, for $\epsilon_i = -1$, we have $g^{\mu\nu}_{i}n_{(i)\mu}(\nabla_{\nu} \rho) = -1 < 0$, which implies that the normal points in the direction of decreasing $\rho$ as seen from the point of view of an observer in $\mathcal{M}_i$. 
 
For $\rho = a$ the metric~\eqref{eq:Dmetric} defined in $\mathcal{M}_i$ reduces to the metric~\eqref{eq:Smetric} defined in $\mathcal{S}$, the projection vectors are $e_{(i)t}^\mu = (1,\, 0,\, 0,\, 0)$ and  $e_{(i)\varphi}^\mu = (0,\,0,\,0,\, 1)$, and then the extrinsic curvature of $\mathcal{S}$ as seen from $\mathcal{M}_i$ is given by
\begin{align}
    K^t_{(i) t} = 0, \ \ K^{t}_{(i) \varphi} =0= \ K^{\varphi}_{(i) t},   \ \ K^{\varphi}_{(i) \varphi} = \frac{\epsilon_i}{a}.  \label{eq:K-}
\end{align}

Obviously, the frame-dragging velocity in the interior region vanishes since the $g_{\varphi t}$ component of the interior metric \eqref{eq:Dmetric} is identically zero.

\subsection{The exterior region}

In the spacetime $(\mathcal{M}_e, g_e)$ exterior to the ring, the metric is \eqref{eq:Ametric} and the spacelike unit normal vector to the boundary $\mathcal{S}$ as seen from $\mathcal{M}_e$ is given by
\begin{equation}
 n_{(e) \mu} = \epsilon_e \delta_{\mu}^{\varrho}, \label{eq:normal+}  
\end{equation}
where $\epsilon_e$ is the normalization factor. The condition $g^{\mu\nu}_en_{(e)\mu}n_{(e)\nu} = 1$ with the metric \eqref{eq:Ametric} implies that $\epsilon_e = \pm 1$ and, therefore, determines the sign of $n_{(e)\mu}$. The sign $\epsilon_e$ is the quantity that determines the relative orientation between $n_{(e)\mu}$ and the coordinate $\varrho$. 
For $\epsilon_e = +1$, we have $g^{\mu\nu}_e n_{(e)\mu}(\nabla_{\nu} \varrho) = +1 > 0$, which implies that the normal points in the direction of increasing $\varrho$ as seen from $\mathcal{M}_e$. In turn, for $\epsilon_i = -1$, we have $g^{\mu\nu}_en_{(e)\mu}(\nabla_{\nu} \varrho) = -1 < 0$, which implies that the normal points in the direction of decreasing $\varrho$ as seen in $\mathcal{M}_e$. Since, for $\varrho = a$, the metric~\eqref{eq:Ametric} of $\mathcal{M}_e$ reduces to the metric~\eqref{eq:Smetric} of $\mathcal{S}$, the projection vectors are identical to the ones for the interior region ${\cal M}_i$, $e_{(i)t}^\mu = (1,\, 0,\, 0,\, 0)$ and  $e_{(i)\varphi}^\mu = (0,\,0,\,0,\, 1)$, then the extrinsic curvature of $\mathcal{S}$ as seen from $\mathcal{M}_e$ is given by
\begin{equation}
 \!\!  K^t_{(e) t} = -2m_0a \epsilon_e, \; K^{t}_{(e) \varphi} =0= \ K^{\varphi}_{(e) t}, \; K^{\varphi}_{(e) \varphi} = \frac{\epsilon_e}{a}. \label{eq:K+}
\end{equation}

As a final comment at this point, we observe that the frame-dragging velocity in the exterior region close to the string vanishes, since the $g_{\varphi t}$ metric component of the exterior metric \eqref{eq:Ametric} is identically zero.

\subsection{Are all thin shell solutions possible?}
\label{sec:thinshell}

\subsubsection{The matter-energy content of the ring}

The Darmois-Israel formalism allows for both values of $\epsilon_i$ and $\epsilon_e$, which, in principle, leads to four possible thin shell solutions. However, the particular choices of $\epsilon_i$ and $\epsilon_e$ determine the possible domains of the coordinates $\rho$ and $\varrho$ that, in the end, must be in accordance with the original domain previously established for the Boyer-Lindquist radial coordinate, $0\leq r<\infty$. 

In order to perform the required analysis, we first describe the fluid properties of the thin shell at the ring $\mathcal{S}$. 
From Eqs.~\eqref{eq:EMT}, \eqref{eq:K-}, and $\eqref{eq:K+}$ it follows that the energy-momentum tensor of the thin shell can be decomposed as 
\begin{equation}
S_{ab} = (\sigma + p) u_a  u_b + p h_{ab}, \label{eq:staticemt0}
\end{equation}
where 
\begin{align}
    \sigma = -S^t_{\ t} = \frac{1}{8 \pi a}(\epsilon_i - \epsilon_e), \ \ p = S^{\varphi}_{\ \varphi} = -\frac{m_0 a }{4 \pi} \epsilon_e,  \label{eq:staticemt}
\end{align}
are the energy density and the pressure (or tension) of the fluid of the ring, respectively, $h_{ab}$ is the metric of the ring given in Eq.~\eqref{eq:Smetric}, and 
\begin{align}
u_a = \delta_a^{\ t},    \label{eq:Unonrot}
\end{align}
is the fluid velocity at the ring.

Notice that the velocity \eqref{eq:Unonrot} does not carry a component along the azimuthal direction $\varphi$, which implies that the string fluid does not rotate with respect to the stationary observer at infinity.  In Ref.~\cite{Spallucci}, the equatorial angular velocity of the G\"urses-G\"ursey fluid in the limit $r = 0$, $\Omega=1/a$ [see Eq.~\eqref{eq:GGveloc}], is attributed to the thin shell itself. However, this assumption is not appropriate once the G\"urses-G\"ursey background and the thin shell have different matter contents, as it can be seen from the different energy-momentum tensors specified by Eqs.~\eqref{eq:ggt} and~\eqref{eq:staticemt}, respectively.

In addition to the vanishing of the fluid angular velocity, as we comment in Sec.~\ref{sec:stmetric}, the frame-dragging of the non-singular Kerr metric vanishes at the ring {$\cal S$} as well, a result that is also consistent with the approximate metric \eqref{eq:Ametric}, so that no rotational effect is observed at the string.

Let us now look at the four possible different cases for the choices of $\epsilon_i$ and $\epsilon_e$.

\subsubsection{The case  $\epsilon_i = +1$ and $\epsilon_e = -1$}
\label{sec:epsi+e-}

Starting with the choice made by Smailagic and Spallucci~\cite{Spallucci}, i.e., $\epsilon_i = +1$ and $\epsilon_e = -1$, we get the line energy density and the pressure of the string as $\sigma = 1/ 4\pi a$ and $p = m_0 a/4\pi$, respectively. 
The line energy density is always positive, while the pressure can be positive or negative depending on the sign of $m_0$.

\begin{figure}[h]
    \centering
    \includegraphics[width=7cm]{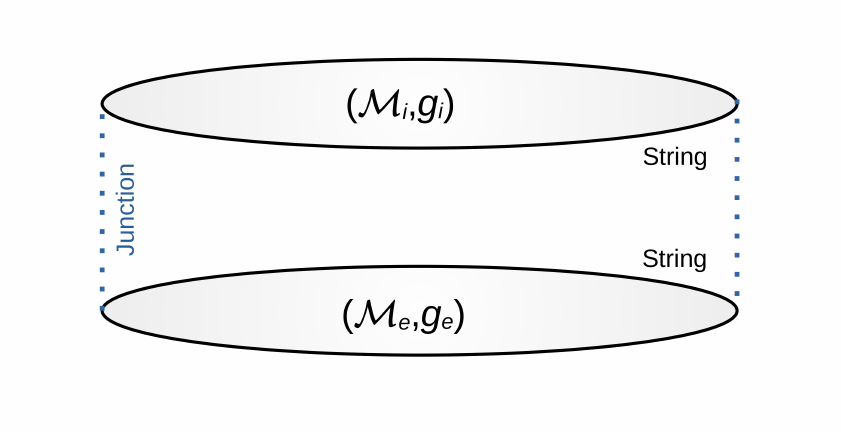}
    \caption{Embedding diagram of a $t = \text{constant}$ and $\theta = \pi/2$ slice of $(\mathcal{M}_i, g_i)$ with $\epsilon_i = +1$ and $(\mathcal{M}_e, g_e)$ with $\epsilon_e = -1$ in $3$-dimensional Euclidean space. The domain of the interior radius is $0 \le \rho \le a$. The domain of the exterior radius is also $0 \le \varrho \le a$. The string is represented by the black solid lines in each spacetime, with the lines being identified through the junction.}
    \label{fig:first}
\end{figure}

More importantly, let us notice that $\epsilon_i = +1$ implies that the normal vector points in the direction of increasing radial coordinate, as seen from the interior region. Since the coordinate $\rho$ is in the interval $\rho \in [0, \, a)$, and recalling our convention that the normal vector $n_\mu$ points from $\mathcal{M}_i$ to $\mathcal{M}_e$, the choice $\epsilon_i = +1$ implies that there is a center in the interior region, as expected. In turn, the choice $\epsilon_e = -1$ implies that the normal vector points in the direction of decreasing radial coordinate, as seen from the exterior region. Since $n_\mu$ points from $\mathcal{M}_i$ to $\mathcal{M}_e$, this implies that the exterior region is also a disk with radius $a$, $\varrho \in [0,\, a]$. Hence, as it can be seen from Fig.~\ref{fig:first}, this possible solution has two centers, one in each region, 
which means that a disk of the Minkowski (interior) geometry is glued to another disk-shaped section of the exterior geometry. The resulting space is a kind of Minkowski-dS or Minkowski-AdS closed universe that can be interpreted as a bubble universe, as is similarly done in Ref.~\cite{Lemos22} for a Minkowski-Minkowski closed universe.
However, the original domain of the coordinate $\varrho$ is $\varrho \in [a, a + \varepsilon)$, for $\varepsilon > 0$, with the already established notion that the exterior region corresponds to the equatorial plane just outside the ring. Therefore, this choice does not lead to the appropriate solution for the thin shell at the boundary $\mathcal{S}$.

\subsubsection{The case  $\epsilon_i = -1$ and $\epsilon_e = +1$} 
\label{sec:epsi-e+}

Another nontrivial case is given by the choice $\epsilon_i = -1$ and $\epsilon_e = +1$. In this case, the line energy density and the pressure of the fluid are given by $\sigma = -1/ 4\pi a$ and $p = -m_0 a/4\pi$, respectively. The line energy density is negative, while the pressure can be positive or negative depending on the sign of $m_0$.

\begin{figure}[t]
    \centering
    \includegraphics[width=7.5cm]{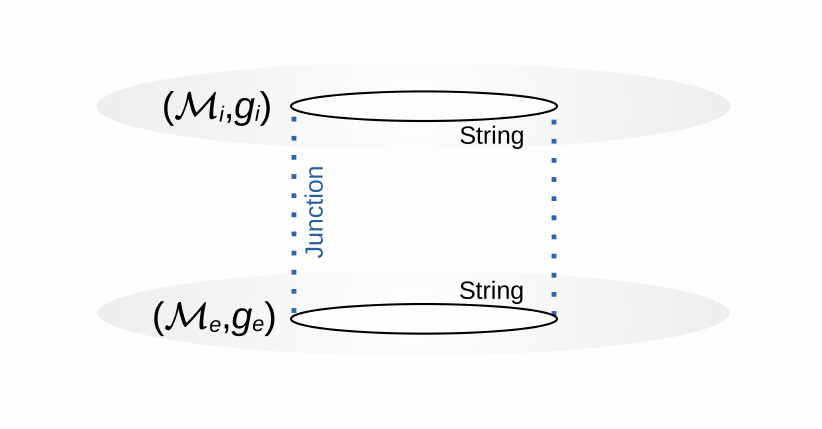}
    \caption{Embedding diagram of a $t = \text{constant}$ and $\theta = \pi/2$ slice of $(\mathcal{M}_i, g_i)$ with $\epsilon_i = -1$ and $(\mathcal{M}_e, g_e)$ with $\epsilon_e = +1$ in $3$-dimensional Euclidean space. The domain of the interior radius is $a \le \rho < \infty$. The domain of the exterior radius is also $a \le \varrho < \infty$. The string is represented by black solid lines in each spacetime, with the lines being identified through the junction.}
    \label{fig:second}
\end{figure}

More importantly, the choice $\epsilon_i = -1$ implies that the normal points in the direction of decreasing radial coordinate as seen from the interior region, while $\epsilon_e = +1$ implies that the normal points in the direction of increasing radial coordinate as seen from the exterior region. Since $n_\mu$ points from $\mathcal{M}_i$ to $\mathcal{M}_e$, this implies that in both regions the radial coordinates start from $a$ and go to infinity. Therefore, the possible solution has no center at all, as it can be seen from Fig.~\ref{fig:second},  
which means that the asymptotic section of the Minkowski (interior) geometry, from which a disk of radius $r = a$ has been removed, is being glued to the asymptotic section of the exterior geometry. The resulting space is a kind of Minkowski-dS or Minkowski-AdS open universe and can be interpreted as a traversable wormhole, as is similarly done in Ref.~\cite{Lemos22} for a Minkowski-Minkowski open universe.
However, once more let us notice that the original domain of the interior radial coordinate is $\rho \in [0,a]$, with the already established notion of the interior region corresponding to the region inside the ring, i.e., corresponding to the disk. Therefore, this choice does not lead to the appropriate solution for the thin shell at the boundary $\mathcal{S}$.

\subsubsection{The case  $\epsilon_i = -1$ and $\epsilon_e = -1$}
\label{sec:epsie-1}

For the choice $\epsilon_i = \epsilon_e = -1$, the line energy density and the pressure of the fluid are given by $\sigma = 0$ and $p = m_0 a/4\pi$, respectively. The line energy density is identically zero, while the pressure can be positive or negative depending on the sign of $m_0$. 

\begin{figure}[t]
    \centering
    \includegraphics[width=7.5cm]{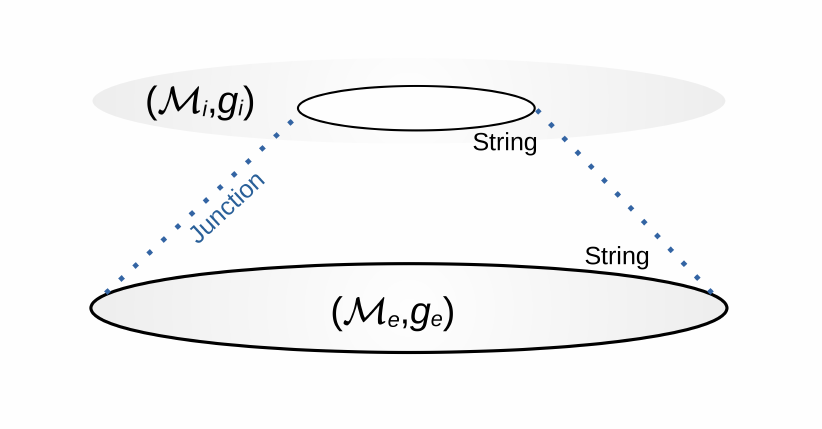}
    \caption{Embedding diagram of a $t = \text{constant}$ and $\theta = \pi/2$ slice of $(\mathcal{M}_i, g_i)$ with $\epsilon_i = -1$ and $(\mathcal{M}_e, g_e)$ with $\epsilon_e = -1$ in $3$-dimensional Euclidean space. The domain of the interior radius is $a \le \rho < \infty$. The domain of the exterior radius is $0 \le \varrho \le a$. The string is represented by the black solid lines in each spacetime, with the lines being identified through the junction.}
    \label{fig:third}
\end{figure}

In this case, the roles of $\mathcal{M}_i$ and $\mathcal{M}_e$ are basically swapped. More importantly, note that this choice for $\epsilon_i$ and $\epsilon_e$ implies that the normal points in the direction of decreasing radial coordinates, as seen from both regions. Since $n_\mu$ points from $\mathcal{M}_i$ to $\mathcal{M}_e$, the role of the radial coordinates $\rho$ and $\varrho$ are swapped. This means that a disk of the de Sitter (exterior) geometry is being glued to the asymptotic section of the Minkowski (interior) geometry from which a disk of radius $r = a$ has been removed, as illustrated in Fig.~\ref{fig:third}. The resulting space is a kind
of dS-Minkowski or AdS-Minkowski universe that presents a central disk that contains a dS(AdS)-type of fluid, and is void of matter everywhere outside the disk. However, this contradicts the original domains of the radial coordinates and the notion that the interior region corresponds to the interior disk and the exterior region corresponds to the equatorial disk just outside the ring. Therefore, this choice does not lead to the appropriate solution for the thin shell at the boundary $\mathcal{S}$.

\subsubsection{The case  $\epsilon_i = +1$ and $\epsilon_e = +1$}

Finally, for the choice $\epsilon_i = \epsilon_e = + 1$, the line energy density and the pressure of the fluid are given by $\sigma = 0$ and $p = -m_0 a/4\pi$. The line energy density is identically zero, while the pressure can be positive or negative depending on the sign of $m_0$.

\begin{figure}[t]
    \centering
    \includegraphics[width=7.5cm]{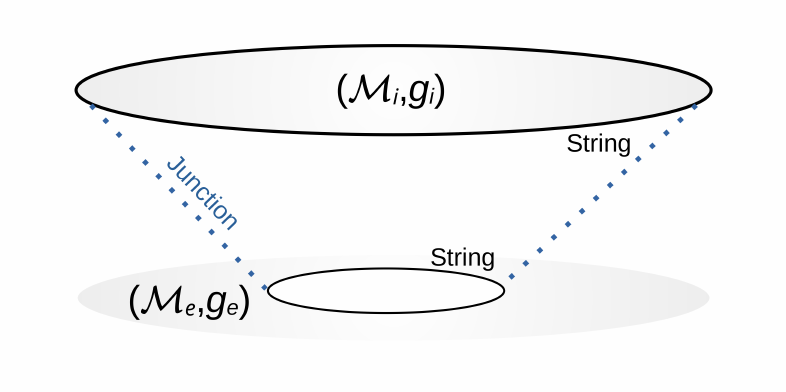}
    \caption{Embedding diagram of a $t = \text{constant}$ and $\theta = \pi/2$ slice of $(\mathcal{M}_i, g_i)$ with $\epsilon_i = +1$ and $(\mathcal{M}_e, g_e)$ with $\epsilon_e = +1$ in $3$-dimensional Euclidean space. The domain of the interior radius is $0 \le \rho \le a$. The domain of the exterior radius is $a \le \varrho < \infty$. The string is represented by the black solid lines in each spacetime, with the lines being identified through the junction.}
    \label{fig:fourth}
\end{figure}

In this case, the features of $\mathcal{M}_i$ and $\mathcal{M}_e$ correspond to the expected notion of the interior and exterior regions. More importantly, let us notice that this choice for $\epsilon_i$ and $\epsilon_e$ implies that the normal points in the direction of increasing radial coordinates, as seen from both regions. Since $n_\mu$ points from $\mathcal{M}_i$ to $\mathcal{M}_e$, the role of the radial coordinates $\rho$ and $\varrho$ stays the same with the domains given by $\rho \in [0,a]$ and $\varrho \in [a, a+\varepsilon)$ with $\varepsilon > 0$, thus respecting the original domain of the radial coordinates. Therefore, this is the choice that corresponds to a consistent solution for the thin shell at $\mathcal{S}$, since it leads to a local geometry that is appropriate to match the global properties of the Kerr geometry with a Minkowski disk glued to a regular dS(AdS) region outside the ring, as illustrated in Fig.~\ref{fig:fourth}. Moreover, the usual assumption in order to construct regular black hole solutions is $m_0 > 0$, from which we notice that the strong energy condition is violated at the ring, which is in agreement with previous results~\cite{Maeda}. 

\section{Delving into the middle-Kerr: The rotating string}
\label{Sec:V}

\subsection{The disk and the ring}

The interior spacetime $(\mathcal{M}_i, g_i)$ is the disk $\mathcal{D}$ with the metric given by Eq.~\eqref{eq:Dmetric}, as described in Sec.~\ref{sec:disk}. Hence, the spacelike unit normal vector to $\mathcal{S}$ as seen from $\mathcal{M}_i$ is the same as in the preceding case, given by \eqref{eq:normal-}, the projection vectors are $e_{(i)t}^\mu = (1,\, 0,\, 0,\, 0)$ and  $e_{(i)\varphi}^\mu = (0,\,0,\,0,\, 1)$, 
and, moreover, the extrinsic curvature of $\mathcal{S}$ as seen from $\mathcal{M}_i$ is also the same as in Eq.~\eqref{eq:K-}.

In turn, the ring $\mathcal{S}$ is described by the same metric given by Eq.~\eqref{eq:Smetric}.

\subsection{The equatorial plane and the exterior metric}

Let us recall that, as the ring is approached from the outside in the equatorial plane, the metric given by Eq.~\eqref{eq:Emetric} reduces to the form \eqref{eq:Ametric0}. In contrast with the metric given by Eq.~\eqref{eq:Ametric} used by the authors in Ref.~\cite{Spallucci}, here we consider the full metric that comes from using $m(\varrho) = m_0 (\varrho^2 - a^2)^{3/2}$ for $\varrho \to a$ in Eq.~\eqref{eq:Emetric}, without any further approximation. In other words, in addition to taking $\theta =\pi/2$ into the full metric, the only approximation we make is choosing the asymptotic form of the mass function $m(\varrho)$. 

Therefore, the exterior spacetime $(\mathcal{M}_e,\,g_e)$ now refers to the equatorial region outside the ring $\mathcal{A}$ with the metric given by Eq.~\eqref{eq:Ametric0}. The spacelike unit normal vector to $\mathcal{S}$ as seen from $\mathcal{M}_e$ is given by \eqref{eq:normal+}, the projection vectors are $e_{(e)t}^\mu = (1,\, 0,\, 0,\, 0)$ and  $e_{(e)\varphi}^\mu = (0,\,0,\,0,\, 1)$, and then the extrinsic curvature of $\mathcal{S}$ is given by
\begin{align}
    & K^t_{(e) t} = -2m_0a \epsilon_e, \ \  K^t_{(e) \varphi} = 2 m_0 a^2 \epsilon_e, \nonumber \\ 
    & K^{\varphi}_{(e) t} = - 2 m_0 \epsilon_e, \ \ K^{\varphi}_{(e) \varphi} = \frac{\epsilon_e}{a} + 2m_0 a \epsilon_e. \label{eq:Kr+}
\end{align}

We now have all the ingredients to perform the matching between the two spacetime regions.

\subsection{The rotating match at the ring}

\subsubsection{{The matter-energy content at the ring}}

As in the preceding case studied in Sec.~\ref{sec:nonrot-ring}, the first boundary condition of the DIF is trivially satisfied. From Eqs.~\eqref{eq:EMT}, \eqref{eq:K-}, and $\eqref{eq:Kr+}$, the second boundary condition gives us
\begin{equation}
\begin{aligned} 
    & 8 \pi S^{t}_{\ t} = 2 m_0 a \epsilon_e + \frac{1}{a}(\epsilon_e - \epsilon_i), \\
    & 8 \pi S^{t}_{\ \varphi} = -  2 m_0 a^2 \epsilon_e, \ \ 
     8 \pi S^{\varphi}_{\ t} =  2 m_0 \epsilon_e,\\
        & 8 \pi S^{\varphi}_{\ \varphi} = -  2 m_0 a \epsilon_e.
\end{aligned} \label{eq:rotstringshell}
\end{equation}
Thus, we can see that, different from the EMT given by Eqs.~\eqref{eq:staticemt0} and \eqref{eq:staticemt}, in this case the EMT of the shell is not diagonal. In fact, the presence of nondiagonal terms such as $\mathcal{S}^t_{\ \varphi}$ is an indication that the fluid at the shell is rotating with respect to the asymptotic observer. 

As we have seen in the previous section, the appropriate interpretation of the energy density and pressure, as well as the resulting topology of the complete spacetime, depends on the particular values of $\epsilon_i$ and $\epsilon_e$, and then it is interesting to study each one of the fourth combinations separately.

\subsubsection{The cases in which $\epsilon_i = - \epsilon_e$}
We first study the interpretation of the EMT \eqref{eq:rotstringshell}by considering $\epsilon_i = - \epsilon_e$. 

The characteristic equation for the matrix~\eqref{eq:rotstringshell} give us $\lambda^2 - 2 \epsilon \lambda /a - 4m_0=0,$  whose solutions are the eigenvalues 
\begin{align}
    \lambda_{\pm} = \frac{\epsilon_e }{8 \pi a}\left(1 \pm \sqrt{1 + 4 m_0 a^2} \right), \label{eq:eigenval}
\end{align}
and the respective normalized eigenvectors are
\begin{align}
     \mathfrak v_\pm^a = N_\pm \left(\frac{1 + 2m_0a^2 \pm \sqrt{1 +4m_0a^2}}{2m_0\,a } \delta_t^a + \delta_\varphi^a\right),
\end{align}
where 
\begin{equation}
N_\pm^{-2} =  \pm\left[a^2 -\left(\frac{1 + 2m_0a^2 \pm \sqrt{1 +4m_0a^2}}{2m_0 a}\right)^2\right] .
\end{equation}
Considering that $m_0$ may assume negative values, we find three different situations for the eigenvalues $L_\pm$ depending on the discriminant $D= 1 + 4 m_0 a^2$: Case {\it (i)} $D > 0$,  case {\it (ii)} $D = 0$, and case {\it (iii)} $D <0 $.  We analyze each case separately.

\vskip .4cm 
\noindent
\textbf{\emph{Case (i)}}: If $1 + 4 m_0 a^2 > 0$, the two eigenvalues $\lambda_{\pm}$ assume real values and the canonical form of the EMT \eqref{eq:rotstringshell} is 
\begin{equation}\label{eq:fluidi-e}
    S_{\ a}^b= \sigma u_a u^b+ p x_a x^b = \left(\sigma+p\right)u_a u^b + p h_a^b,
\end{equation}
where 
\begin{equation}\begin{split}
& \sigma = -\lambda_+,\qquad u^a = v_+ ^a \\
& p = \lambda_-,\; \;\  \qquad x^a=  v_-^a, 
\end{split}
\end{equation}
with $u^a$ being a timelike vector and $x^a$ being a spacelike vector.

Moreover, the timelike eigenvector $u^a= v^{a}_{+}$ can be decomposed in terms of the coordinate basis as
\begin{align}
    u^a = \gamma\left( \delta^{a}_{\ t} + \omega \delta^{a}_{\ \varphi}\right), 
\end{align}
where $\gamma =\left(1-\omega^2 a^2\right)^{-1/2}$ and 
\begin{align}
    \omega & =  \frac{2m_0a}{1+ 2m_0 a^2 +\sqrt{1 + 4 m_0 a^2}}.\label{eq:omega}
\end{align}
can be interpreted as the angular velocity of the matter fluid along the rotating string.

The stress-energy tensor \eqref{eq:fluidi-e} can be interpreted as a perfect fluid. 
For $\epsilon_e = -1$, which also means $\epsilon_i=+1$, the energy density of the shell $\sigma$ is positive, while the pressure $p$ is negative. 
For $\epsilon_e = +1$, which also means $\epsilon_i=-1$, the energy density of the shell $\sigma$ is negative, while the pressure $p$ is positive. 
 
\vskip .4cm 
\noindent
\textbf{\emph{Case (ii)}}: If $1 + 4 m_0 a^2 = 0$, the two eigenvalues $\lambda_{\pm}$ are degenerated (identical) and the canonical form of the EMT \eqref{eq:rotstringshell} is 
\begin{equation}\label{eq:fluidi-e2}
 S_{\ a}^b=\sigma \left(u_a u^b + x_a x^b\right) + \beta k_a k^b, 
\end{equation}
where 
\begin{equation}\begin{split}
& \sigma = -p =-\frac{\epsilon_e}{8\pi a}, \qquad \beta= -\frac{\epsilon_e}{16 \pi a},\\
& k^a =   \delta_{\ t}^a - \frac{1}{a}\delta_{\ \varphi}^a,\  \qquad 
 u^a =  \delta_{\ t}^a, \qquad x^a= \frac{1}{a}\delta_{\ \varphi}^a,
\end{split}
\end{equation}
with $k^a$ being a lightlike vector, $u^a$ being a timelike vector, and $x^a$ being a spacelike vector. 

The stress-energy tensor \eqref{eq:fluidi-e2} can be interpreted as a mixture of a perfect fluid with a flux of directed radiation (lightlike fluid) along the string. 
For $\epsilon_e = -1$, which also means $\epsilon_i=+1$, the energy density of the matter fluid $\sigma$ and the energy density of the lightlike fluid $\beta$ are both positive, while the pressure $p =-\sigma$ of the matter fluid is negative. For $\epsilon_e = +1$, which also means $\epsilon_i=-1$, the energy density of the matter fluid $\sigma$ and the energy density of the lightlike fluid $\beta$ are both negative, while the pressure $p=-\sigma$ of the matter fluid is positive.  

\vskip .4cm 
\noindent
\textbf{\emph{Case (iii)}}: If $1 + 4 m_0 a^2 < 0$, the two eigenvalues $\lambda_{\pm}$ assume complex values, and the canonical form of the EMT \eqref{eq:rotstringshell} is 
\begin{equation}\label{eq:fluidi-e3}
    S^a_{\ b} =  \sigma \left(u^a u_b - x^a x_b\right)  +\beta\left(u^a x_b + x^a u_b\right),
\end{equation}
where 
\begin{equation}\begin{split}
& \sigma = -\frac{\epsilon_e}{8 \pi\, a}, \qquad  p = -\sigma=  \frac{\epsilon_e}{8 \pi\, a},  \\
& \beta = -\frac{\epsilon_e}{8\pi\, a}\sqrt{|1+4m_0a^2|}, \\
& u^a = \sqrt{\frac{1}{2} + \Big|\frac{m_0 a}{8 \pi \beta}\Big|}\,\delta_{\ t}^a +  \frac{1}{a}\sqrt{\Big|\frac{m_0 a}{8 \pi \beta}\Big| -\frac12}\, \delta_{\ \varphi}^a \\
& x^a = \sqrt{\Big|\frac{m_0 a}{8 \pi \beta}\Big| -\frac12}\,\delta_{\ t}^a + \frac{1}{a}\sqrt{\frac{1}{2} + \Big|\frac{m_0 a}{8 \pi \beta}\Big|}\,\delta_{\ \varphi}^a   , 
\end{split}
\end{equation}
with $u^a$ being a unit timelike vector and $x^a$ being a unit spacelike vector.

The stress-energy tensor \eqref{eq:fluidi-e3} can be interpreted as a (perfect) fluid with heat flow along the string, with $u^a$ representing the four-velocity of the fluid and $x^a$ giving the direction of the heat flow. For $\epsilon_e = -1$, which also means $\epsilon_i=+1$, the energy density of the shell $\sigma$ is positive, while the pressure $p$ is negative. Since $\beta>0$, the heat flux is along $x^a$ spacelike direction.
For $\epsilon_e = +1$, which also means $\epsilon_i=-1$, the energy density of the shell $\sigma$  is negative, while the pressure $p$ is positive. The change from $\beta> 0$ to $\beta< 0$ indicates the reversion the heat flux with respect to $x^a$.

Now we consider the topology of the resulting spacetime in the two cases engendered by the choice  $\epsilon_i = - \epsilon_e$. 
For $\epsilon_e = -1$, which also means $\epsilon_i=+1$, the situation is identical to case discussed in Sec.~\ref{sec:epsi+e-}, whose resulting topology is depicted in Fig.~\ref{fig:first}.  This scenario represents a closed universe with a rotating string at the ring joining the two regions.
For $\epsilon_e = -1$, which also means $\epsilon_i=+1$, the situation is identical to case discussed in Sec.~\ref{sec:epsi-e+}, whose resulting topology is depicted in Fig.~\ref{fig:second}. This scenario corresponds to an open universe with a rotating string at the ring joining the two regions.

\subsubsection{The cases in which $\epsilon_i = \epsilon_e$}

If $\epsilon_i = \epsilon_e$, it is not possible to diagonalize the EMT of the shell since the determinant of $S^{a}_{\ b}$ vanishes and, therefore, the EMT can be interpreted as a lightlike dust fluid. This is indeed the case once that $S^{a}_{\ b}$ may be recast as
\begin{align}
    S^{a}_{\ b} = H k^a k_b,
\end{align}
where $k^a$ is a lightlike vector given by
\begin{align}
    k^a = \delta^{a}_{\ t} + \frac{1}{a} \delta^{a}_{\ \varphi},
\end{align}
with $\delta^{a}_{b}$ standing for the Kronecker delta. The quantity $H$ is given by
\begin{align}
    H = - \frac{m_0 a \epsilon_e}{4 \pi}. 
\end{align}
In fact, in the present case, $H$ is a constant and can be interpreted as the energy density of the shell. Hence, the thin-shell ring corresponds to a string composed of a rotating lightlike dust that rotates with angular velocity $1/a$.

In the case with $\epsilon_e = -1$, which also means $\epsilon_i =-1$, the energy density of the string depends on the sign of $m_0$, which is positive for $m_0>0$.  Similarly to the case discussed in Sec.~\ref{sec:epsie-1}, the roles of $\mathcal{M}_i$ and $\mathcal{M}_e$ are swapped, and the resulting geometry does not correspond to the original problem. See also Fig.~\ref{fig:third}.

In the case with $\epsilon_e= 1$, which also means $\epsilon_i =1$, the energy density of the string depends on the sign of $m_0$, which is positive for $m_0<0$. Moreover, $\mathcal{M}_i$ and $\mathcal{M}_e$ correspond to the expected notion of interior and exterior regions, respectively, with both regions being joined by a rotating string at the ring (see also Fig.~\ref{fig:fourth}). 
Hence, this is an appropriate choice to match the overall geometry of the Kerr-like spacetime with a regular rotating core. However, there is a drawback regarding the kind of matter of the string. The energy density of the string is negative for $m_0 >0$, while the center core of the Kerr-like geometry presents a rotating de Sitter-type fluid with a positive energy density. Interestingly, this case is similar to the string source investigated by Israel in Ref.~\cite{Israel77} for the Kerr geometry,in which the ring is composed of a rotating dust-like material that also rotates with angular velocity $1/a$. In such a work by Israel, the energy density is negative in the region $r < 0$. Here, we do not investigate the extension through the region $r < 0$ for the Kerr-like geometry. On the other hand, the choice $m_0 <0$ leads to a string with a positive energy density and the center core of the Kerr-like geometry contains a rotating anti-de Sitter-type fluid.

\section{Discussion}
\label{sec:conc}

We studied the central region of regular Kerr-like geometries with a variable mass function that depends on the radial Boyer-Lindquist radius as $m(r) \simeq r^3$.  First, we reviewed the behavior of the main curvature scalars close to $r=0$ and confirmed that such scalars present a jump at the Kerr ring. Then, following the proposal by Smailagic and Spallucci \cite{Spallucci}, in which the problematic central ring in the Kerr-like geometries is replaced with a string, whose interior region is flat and whose exterior region is described by a de Sitter-type metric, we studied two different cases.

The first case analyzed is the same spacetime construction studied in \cite{Spallucci}, but here considering all possible thin shell configurations that can be built from the Darmois-Israel formalism, regarding the possible orientations of the normal vector relative to the radial coordinate. The different orientations of the normal vector engender four different configurations and can be parameterized by two unit parameters $\epsilon_e=\pm1$ and $ \epsilon_e=\pm 1$.

Considering the original choice $\epsilon_i = -\epsilon_e = +1$ made by Smailagic and Spallucci~\cite{Spallucci}, we conclude that the resulting solution is not appropriate for a thin shell at the ring of a rotating regular black hole, once the resulting spacetime topology contradicts the original domain of the exterior radial coordinate and the notion of the exterior region as the equatorial region just outside the ring. In fact, such a choice gives rise to a closed spacetime that may be interpreted as a Minkowski-de Sitter closed universe, corresponding to a bubble universe. More importantly, we formally demonstrated that the only consistent solution for the thin shell at the ring of a rotating regular black hole results from the choice $\epsilon_i = \epsilon_e = +1$. In this case, the original domains of the radial coordinates are respected and the roles of the interior and exterior regions are preserved. Hence, differently from Ref.~\cite{Spallucci}, we find that the suitable string solution replacing the Kerr ring has a vanishing line energy density and a pressure that can be either positive or negative depending on the sign of the constant $m_0$. 

Finally, we noticed that in the scenario considered in Ref.~\cite{Spallucci}, the string of matter replacing the ring is not rotating. This result is obtained by verifying that both the angular velocity of the matter and the frame-dragging effect in the string are vanishing.

The second case analyzed here is a spacetime construction similar to that of the first case, but now the exterior metric is different from that considered in \cite{Spallucci}. More specifically, we constructed a new scenario in which the string is indeed rotating and considered the four possible combinations of normal vector orientations. The results for the corresponding geometries are the same as for the nonrotating string, while the kinds of material result quite different. In cases where $\epsilon_i = - \epsilon_e$, the matter content of the rotating string corresponds to a fluid with a positive energy density for $\epsilon_i = -\epsilon_e= -1$, and with a negative energy density for $\epsilon_i=- \epsilon_e = +1$. In the case where $\epsilon_i = \epsilon_e$,  the rotating string corresponds to a lightlike fluid with a positive energy density for $m_0\epsilon_e <0$, and with a negative energy density for $m_0\epsilon_e >0$. Since the combination that leads to an appropriate solution with respect to the resulting spacetime topology is $\epsilon_e = \epsilon_i =1$, we have two possible Kerr-like solutions that depends on the sign of $m_0$. For $m_0 < 0$, the energy density of the string is positive, which means a negative mass function for the regular Kerr geometry and is interpreted as an anti-de Sitter-type of matter. For $m_ 0 > 0$, the energy density of the string is negative, however the mass function of the regular Kerr geometry is positive and is interpreted as a de Sitter-type of matter.

\begin{acknowledgments}

M.~L.~W.~B.~is funded by Funda\c c\~ao de Amparo \`a Pesquisa do Estado de S\~ao Paulo (FAPESP), Brazil, Grant No.~2022/09496-8. V.~T.~Z.~thanks partial financial support from Conselho Nacional de Desenvolvimento Cien\-t\'ifico
e Tecnol\'ogico (CNPq), Brazil, Grant No.~311726/2022-4, and from Funda\c c\~ao de Aperfei\c coa\-men\-to do Pessoal de N\'ivel Superior (CAPES), Brazil, Grant No. 88887.310351/2018-00.

\end{acknowledgments}


\end{document}